# Servo-Controllers with Operational Constraints

Eugene Lavretsky

*Abstract* – In this paper, a proportional-integral servo-control design method is developed for multi-input-multi-output linear time invariant systems with operational constraints imposed on the system control input and on an output of the same dimension as the control input. The design is based on min-norm controllers and Control Barrier Functions. It allows to enforce min/max box constraints by analytically solving Quadratic Programs for min-norm augmentation controllers. The method provides an anti-windup protection for the controller integrator state and enforces the desired operational control and output constraints, component-wise. A simulation example is given to illustrate potential benefits of the proposed design methodology for aerial flight critical systems.

## 1. Introduction and Problem Formulation

Consider the Multi-Input-Multi-Output (MIMO) Linear Time Invariant (LTI) dynamical system with control position constraints,

$$\text{Open-Loop System Dynamics}: \dot{x}_p = A_p x_p + B_p \underbrace{\text{sat}_{u^{\min}}^{u^{\max}}\left(u^{cmd}\right)}_{u}, \quad x_p \in R^{n_p}$$

$$\text{Regulated Output}: y_{reg} = C_{p\,reg} x_p + D_{p\,reg} u, \quad y_{reg} \in R^m \quad (1.1)$$

$$\text{Limited Output}: z_{\lim} = C_p x_p + D_p u, \quad z_{\lim} \in R^m$$

where $x_p \in R^{n_p}$ is the $n_p$ – dimensional state vector, $(u, u_{cmd}) \in R^m$ are the $m$ – dimensional commanded and the achieved control inputs, $\text{sat}_{u^{\min}}^{u^{\max}}(*)$ is the static saturation function with component-wise min/max bounds $(u^{\min}, u^{\max}) \in R^m$ imposed on $u^{cmd}$, $y_{reg} \in R^m$ is the system $m$ –dimensional vector of regulated outputs, and $z_{\lim} \in R^m$ is the $m$ – dimensional limited output to be kept within the desired min/max bounds $(z_{\lim}^{\min}, z_{\lim}^{\max}) \in R^m$, component-wise.

Generalizations to $z_{\lim} \in R^p$ with $m < p \le n_p$ are possible but will not be considered within this paper for clarity sake. Also, it is assumed that the relative degree of $z_{\lim}$ is either zero or one, that is either $\det(D_p) \ne 0$ or $\det(C_p B_p) \ne 0$, respectively. The proposed in this paper control design can be extended to outputs with higher relative degrees. That extension will be reported at a later time.

In (1.1), the system matrices $(A_p, B_p, C_p, D_p, C_{p\,reg}, D_{p\,reg})$ are of the corresponding dimensions, $A_p$ is Hurwitz, and the matrix pair $(A_p, B_p)$ is controllable. It is further assumed that the entire state vector $x_p$ is accessible for control design, as the system output measurement.

Of interest is the servo-control design problem with min/max (called "box") operational constraints. Specifically, a state feedback commanded control input $u^{cmd}$ needs to be found such that the system regulated output $y_{reg}$ tracks external commands $y_{cmd} \in R^m$, while the control command $u^{cmd}$ and the limited output $z_{\lim}$ evolve within their predefined min/max operational constraint bounds, component-wise.

$$u^{\min} \le u^{cmd} \le u^{\max}, \quad z_{\lim}^{\min} \le z_{\lim} \le z_{\lim}^{\max} \quad (1.2)$$

If such a servo-controller can be designed than the saturation (sat) function in the system dynamics can be removed. In that sense, operational constraints (1.2) are often called "soft" to distinguish them from the

"hard" constraints that are enforced and represented by the sat function. This paper derives a servo-controller with soft operational constraints (1.2).

In order to facilitate robust tracking of external commands while operating in the presence of limits (1.2), consider the integrated output tracking error dynamics with the anti-windup (AW) control modification term $v \in R^m$,

$$\dot{e}_{yI} = y_{reg} - y_{cmd} + v \tag{1.3}$$

which is to be designed such that the integrator state $e_{yI} \in R^m$ is uniformly bounded during control saturation events [1], [2].

If a component of the commanded control $u^{cmd}$ exceeds its min/max bounds then that control channel is saturated at the corresponding limit and the system becomes open-loop with respect to that control input component. During control saturation events, the controller integrator state $e_{yI}$ needs to be kept bounded irrespective of the tracking error dynamics, which in turn can drive the integrator state to become unbounded, that is it would "wind-up" [1], [2]. The AW control modification input $v$ in (1.3) will be designed to prevent the integrator state from winding up.

Augmenting (1.1) with (1.3), gives the $n = (n_p + m)$ – dimensional extended open-loop system.

$$\underbrace{\begin{pmatrix} \dot{e}_{yI} \\ \dot{x}_p \end{pmatrix}}_{\dot{x}} = \underbrace{\begin{pmatrix} 0_{m \times m} & C_{p\,reg} \\ 0_{n_p \times m} & A_p \end{pmatrix}}_{A} \underbrace{\begin{pmatrix} e_{yI} \\ x_p \end{pmatrix}}_{x} + \underbrace{\begin{pmatrix} D_{p\,reg} \\ B_p \end{pmatrix}}_{B} u + \underbrace{\begin{pmatrix} -I_m \\ 0 \end{pmatrix}}_{B_{cmd}} (y_{cmd} - v) \tag{1.4}$$

It can be proven that the extended pair $(A, B)$ is controllable if and only if $\begin{pmatrix} A_p & B_p \\ C_{p\,reg} & D_{p\,reg} \end{pmatrix}$ is nonsingular. That is equivalent to require that the original system (1.1) with the regulated output $y_{reg}$ has no transmission zeros at the origin, which is assumed to take place going forward.

For the unconstrained servo-control problem, a baseline state feedback control policy,

$$u_{bl}^{cmd} = -\underbrace{K_x}_{(K_I\ K_P)} \underbrace{x}_{\begin{pmatrix} e_{yI} \\ x_p \end{pmatrix}} = \underbrace{-K_I e_{yI}}_{u_I} \underbrace{-K_P x_p}_{u_P} = u_I + u_P \tag{1.5}$$

can be found such that the system regulated output $y_{reg}$ tracks external constant bounded commands $y_{cmd}$ with zero tracking errors and other bounded commands with sufficiently small errors [3].

In (1.8), $K_I \in R^{m \times m}$ and $K_P \in R^{m \times n_p}$ denote the integral and the proportional feedback gain matrices, respectively, while $u_I$ and $u_P$ represent the integral and state-proportional feedback components of the Proportional-Integral (PI) servo-controller (1.5). Feedback gains for an unconstrained servo-controller such as (1.5) can be computed using the pole placement or the Linear Quadratic Regulator (LQR) control-theoretic design methods [3].

In order to account for the operational constraints (1.2), the total control command is defined as,

$$u^{cmd} = u_{bl}^{cmd} + w \tag{1.6}$$

where $w \in R^m$ is the baseline control augmentation command.

Suppose that a baseline servo-controller (1.5) is designed to yield closed-loop stability and an adequate command tracking performance, without specifically accounting for the operational constraints (1.2). In this case, applying the unlimited servo-controller (1.6) to (1.4), gives the corresponding extended closed-loop dynamics.

$$\begin{aligned}\dot{x} &= Ax + B\left(u_{bl}^{cmd} + w\right) + B_{cmd}\left(y_{cmd} - v\right) \\ &= (A - BK_x)x + Bw + B_{cmd}\left(y_{cmd} - v\right)\end{aligned} \quad (1.7)$$

Adding explicit control limits to (1.6), leads to the modified servo-controller dynamics,

$$\dot{e}_{yI} = y_{reg} - y_{cmd} + v$$

$$u = \operatorname{sat}_{u^{min}}^{u^{max}}\left(\underbrace{-K_I e_{yI}}_{u_I} \underbrace{-K_P x_p}_{u_P} + w\right) = \operatorname{sat}_{u^{min}}^{u^{max}}\left(\underbrace{\underbrace{u_I + u_P}_{u_{bl}^{cmd}} + w}_{u^{cmd}}\right) \quad (1.8)$$

with the AW control modification input $v$ and the control augmentation input $w$. These two control signals will be designed to solve the servo-control problem with the operational constraints (1.2).

## 2. Constrained Quadratic Program for Servo-Control Augmentation Design

Motivated by the min-norm controller design method [4], consider the following Quadratic Program (QP) [5], with box constraints (1.2).

$$\begin{aligned}\text{Minimizaion Cost} &: J(v, w) = \left(v^T R_v v + w^T R_w w\right) \to \min_{v, w} \\ \text{Control Constraints} &: u^{min} \leq \underbrace{\left(u_{bl}^{cmd} + w\right)}_{u^{cmd}} \leq u^{max} \\ \text{Output Constraints} &: z_{lim}^{min} \leq \underbrace{\left(C_p x_p + D_p u\right)}_{z_{lim}} \leq z_{lim}^{max}\end{aligned} \quad (2.1)$$

Let $w_f \in R^m$ denote a filtered version of the control augmentation signal $w$,

$$\dot{w}_f = K_w (w - w_f) \quad (2.2)$$

with a Hurwitz matrix $(-K_w)$. The filter (2.2) provides an approximation to $\dot{w}$. Consider modified control constraints,

$$u^{min} \leq \left(u_{bl}^{cmd} + w_f\right) \leq u^{max} \quad (2.3)$$

and rewrite the QP (2.1) with box constraints in standard form, with single-sided inequality constraints and the modified control limits (2.3).

$$\begin{aligned}\text{Minimizaion Cost} &: J(v, w) = \left(v^T R_v v + w^T R_w w\right) \to \min_{v, w} \\ \text{Control Constraints} &: g\left(u_{bl}^{cmd}\right) = \begin{pmatrix} g_1\left(u_{bl}^{cmd}\right) \\ g_2\left(u_{bl}^{cmd}\right) \end{pmatrix} = \begin{pmatrix} u^{min} - u_{bl}^{cmd} - w_f \\ u_{bl}^{cmd} - u^{max} + w_f \end{pmatrix} \leq 0 \\ \text{Output Constraints} &: h(x_p) = \begin{pmatrix} h_1(x_p) \\ h_2(x_p) \end{pmatrix} = \begin{pmatrix} z_{lim}^{min} - z_{lim} \\ z_{lim} - z_{lim}^{max} \end{pmatrix} \leq 0\end{aligned} \quad (2.4)$$

Note that $w_f$ is used only in (2.4), while the total control command (1.6) remains the same. The main purpose of introducing the filtered control augmentation signal $w_f$ is to allow differentiation of the control constraints and writing the result as a realizable linear combination of the system inputs and outputs, which is discussed next.

In order to solve the QP (2.4), the minimization constraints must be written explicitly in terms of the control decision variables $v$ and $w$. That can be accomplished by differentiating the modified control constraints along the closed-loop system trajectories (1.7), while using (2.2) and regrouping alike terms to collocate the control decision variables.

$$
\begin{aligned}
\dot{g}(x,v,w,w_f) &= \begin{pmatrix} -I_m \\ I_m \end{pmatrix}\left(-K_I(e_y+v) - K_P(A_p x_p + B_p u_{bl}^{cmd} + B_p w) + \boxed{\dot{w}_f}\right) \\
&= \begin{pmatrix} I_m \\ -I_m \end{pmatrix}\left[\underbrace{K_I}_{G_v} v + \underbrace{(K_P B_p - K_w)}_{G_w} w\right] + \underbrace{\begin{pmatrix} I_m \\ -I_m \end{pmatrix}\left(K_I e_y + K_P(A_p x_p + B_p u_{bl}^{cmd}) + K_w w_f\right)}_{\Delta \dot{g}(x,w_f)} \\
&= \begin{pmatrix} I_m \\ -I_m \end{pmatrix}(G_v v + G_w w) + \Delta \dot{g}(x,w_f)
\end{aligned}
\tag{2.5}
$$

The QP control constraints can now be replaced with,

$$
\begin{aligned}
G(x,v,w,w_f) &= \dot{g}(x,v,w,w_f) + \alpha_v g(u_{bl}^{cmd}) \\
&= \begin{pmatrix} G_1(x,v,w,w_f) \\ G_2(x,v,w,w_f) \end{pmatrix} = \begin{pmatrix} I_m \\ -I_m \end{pmatrix}(G_v v + G_w w) + \underbrace{\left[\Delta \dot{g}((x,w_f)) + \alpha_v g(u_{bl}^{cmd})\right]}_{\Delta G(x,w_f)} \\
&= \begin{pmatrix} I_m \\ -I_m \end{pmatrix}(G_v v + G_w w) + \Delta G(x,w_f) \leq 0
\end{aligned}
\tag{2.6}
$$

where

$$
\begin{aligned}
G_v &= K_I \\
G_w &= K_P B_p - K_w \\
\Delta \dot{g}(x,w_f) &= \begin{pmatrix} I_m \\ -I_m \end{pmatrix}\left(K_I e_y + K_P(A_p x_p + B_p u_{bl}^{cmd}) + K_w w_f\right) \\
\Delta G(x,w_f) &= \Delta \dot{g}((x,w_f)) + \alpha_v g(u_{bl}^{cmd})
\end{aligned}
\tag{2.7}
$$

Expression (2.6) represents a linear in control decision variables function with a positive constant $\alpha_v$. A rationale for adding the term $\alpha_v g(u_{bl}^{cmd})$ to the control constraint derivative $\dot{g}(x,v,w,w_f)$ comes from the Nagumo Theorem [6] and the method of Control Barrier Functions (CBF) [7], [8], [9]. It is interesting to note that in [4], this term is called the "negative margin", indicating that its purpose is to repel the system trajectories near their designated limit boundaries.

Returning to the output constraints in (2.4), consider two cases: 1) The limited output $z_{\lim}$ has relative degree zero, $\det(D) \neq 0$; or 2) it is of relative degree one, that is: $D = 0$ and $\det(C_p B_p) \neq 0$.

If $z_{\lim}$ has relative degree zero then no output differentiation is needed and the output constraints can be written directly as linear functions of the control decision variable $w$,

$$
\begin{aligned}
H(x,w) &= \begin{pmatrix} H_1(x,w) \\ H_2(x,w) \end{pmatrix} = \begin{pmatrix} -I_m \\ I_m \end{pmatrix}(C_p x_p + D_p(u_{bl}^{cmd} + w)) + \begin{pmatrix} z_{\lim}^{\min} \\ -z_{\lim}^{\max} \end{pmatrix} \\
&= \underbrace{\begin{pmatrix} -I_m \\ I_m \end{pmatrix} D_p}_{H_w} w + \underbrace{\begin{pmatrix} -I_m \\ I_m \end{pmatrix}(C_p x_p + D_p u_{bl}^{cmd}) + \begin{pmatrix} z_{\lim}^{\min} \\ -z_{\lim}^{\max} \end{pmatrix}}_{\Delta H(x)} = \begin{pmatrix} -I_m \\ I_m \end{pmatrix} H_w w + \Delta H(x) \leq 0
\end{aligned}
\tag{2.8}
$$

where

$$H_w = D_p$$
$$\Delta H(x) = \begin{pmatrix} -I_m \\ I_m \end{pmatrix}\left(C_p x_p + D_p u_{bl}^{cmd}\right) + \begin{pmatrix} z_{\lim}^{\min} \\ -z_{\lim}^{\max} \end{pmatrix} \tag{2.9}$$

If on the other hand, $z_{\lim}$ is of relative degree one then differentiation of the output along the closed-loop system (1.7) trajectories, gives

$$\dot{h}(x,w) = \begin{pmatrix} \dot{h}_1(x,w) \\ \dot{h}_2(x,w) \end{pmatrix} = \begin{pmatrix} -I_m \\ I_m \end{pmatrix} C_p \left( A_p x_p + B_p u_{bl}^{cmd} + B_p w \right) \tag{2.10}$$

and in this case, the output constraints can be replaced with a CBF-based expression, using a positive constant $\alpha_w$,

$$H(x,w) = \dot{h}(x,w) + \alpha_w h(x_p)$$
$$= \begin{pmatrix} -I_m \\ I_m \end{pmatrix} C_p \left( A_p x_p + B_p u_{bl}^{cmd} + B_p w \right) + \alpha_w h(x_p) \tag{2.11}$$
$$= \underbrace{\begin{pmatrix} -I_m \\ I_m \end{pmatrix}(C_p B_p)}_{H_w} w + \underbrace{\begin{pmatrix} -I_m \\ I_m \end{pmatrix} C_p \left( A_p x_p + B_p u_{bl}^{cmd} \right) + \alpha_w h(x_p)}_{\Delta H(x)} = \begin{pmatrix} -I_m \\ I_m \end{pmatrix} H_w w + \Delta H(x) \leq 0$$

where

$$H_w = C_p B_p$$
$$\Delta H(x) = \begin{pmatrix} -I_m \\ I_m \end{pmatrix} C_p \left( A_p x_p + B_p u_{bl}^{cmd} \right) + \alpha_w h(x_p) \tag{2.12}$$

Combining (2.8) and (2.11), the output constraints are written in a generic form,

$$H(x,w) = \begin{pmatrix} -I_m \\ I_m \end{pmatrix} H_w w + \Delta H(x) \leq 0 \tag{2.13}$$

where

$$\text{Output is Relative Degree Zero}: \begin{cases} H_w = D_p \\ \Delta H(x) = \begin{pmatrix} -I_m \\ I_m \end{pmatrix}\left(C_p x_p + D_p u_{bl}^{cmd}\right) + \begin{pmatrix} z_{\lim}^{\min} \\ -z_{\lim}^{\max} \end{pmatrix} \end{cases}$$
$$\text{Output is Relative Degree One}: \begin{cases} H_w = C_p B_p \\ \Delta H(x) = \begin{pmatrix} -I_m \\ I_m \end{pmatrix} C_p \left( A_p x_p + B_p u_{bl}^{cmd} \right) + \alpha_w h(x_p) \end{cases} \tag{2.14}$$

Finally, using the modified CBF-based input-output constraints (2.6), (2.13), results in the following QP formulation.

$$\text{Minimization Cost}: J(v,w) = \left(v^T R_v v + w^T R_w w\right) \to \min_{v,w}$$
$$\text{Control Constraints}: G(x,v,w,w_f) = \begin{pmatrix} I_m \\ -I_m \end{pmatrix}(G_v v + G_w w) + \Delta G(x,w_f) \leq 0 \tag{2.15}$$
$$\text{Output Constraints}: H(x,w) = \begin{pmatrix} -I_m \\ I_m \end{pmatrix} H_w w + \Delta H(x) \leq 0$$

Since the minimization cost and the constraint functions are convex, this QP has the unique optimal solution strategy pair $(v^*, w^*)$, for any given set of external signals $x$ and $w_f$, [5].

## 3. QP-Based Servo-Control Augmentation Design

Given the CBF-based QP formulation in (2.15), consider the corresponding Lagrangian function,

$$L(x,v,w,w_f,\lambda,\gamma) = J(v,w) + \lambda^T G(x,v,w,w_f) + \gamma^T H(x,w)$$
$$= v^T R_v v + w^T R_w w + \lambda^T \left( \begin{pmatrix} I_m \\ -I_m \end{pmatrix} (G_v v + G_w w) + \Delta G(x,w_f) \right) + \gamma^T \left( \begin{pmatrix} -I_m \\ I_m \end{pmatrix} H_w w + \Delta H(x) \right) \quad (3.1)$$

with two Lagrange multiplier vector-coefficients.

$$\lambda = \begin{pmatrix} \lambda_1 \\ \lambda_2 \end{pmatrix} \in R^{2m}, \quad \lambda_k \in R^m, \quad k = 1,2$$
$$\gamma = \begin{pmatrix} \gamma_1 \\ \gamma_2 \end{pmatrix} \in R^{2m}, \quad \gamma_k \in R^m, \quad k = 1,2 \quad (3.2)$$

With respect to the control decision variables $v$ and $w$, the Lagrangian (3.1) is convex and differentiable. Therefore, Karush-Kuhn-Tucker (KKT) conditions for optimality are applicable for any $x \in R^n$, $v \in R^m$, $w \in R^m$ and $w_f \in R^m$, [5].

$$\begin{aligned}
&\text{Stationarity}: \frac{\partial L(x,v,w,w_f,\lambda,\gamma)}{\partial v} = 0, \quad \frac{\partial L(x,v,w,w_f,\lambda,\gamma)}{\partial w} = 0 \\
&\text{Primal Feasibility}: G(x,v,w,w_f) \leq 0, \quad H(x,w) \leq 0 \\
&\text{Dual Feasibility}: \lambda \geq 0, \quad \gamma \geq 0 \\
&\text{Complementary Slackness}: \begin{cases} \lambda_i G_i(x,v,w,w_f) = 0 \\ \gamma_i H_i(x,w) = 0 \end{cases}, \quad i = 1,2,\ldots,(2m)
\end{aligned} \quad (3.3)$$

Solving the KKT stationarity conditions,

$$\frac{\partial L(x,v,w,w_f,\lambda,\gamma)}{\partial v} = 2 R_v v + G_v^T (I_m \ -I_m) \lambda = 0$$
$$\frac{\partial L(x,v,w,w_f,\lambda,\gamma)}{\partial w} = 2 R_w w + G_w^T (I_m \ -I_m) \lambda - H_w^T (I_m \ -I_m) \gamma = 0 \quad (3.4)$$

for the optimal decision policies $v^*$ and $w^*$, gives

$$v^* = -0.5 R_v^{-1} G_v^T (\lambda_1 - \lambda_2)$$
$$w^* = -0.5 R_w^{-1} \left( G_w^T (\lambda_1 - \lambda_2) - H_w^T (\gamma_1 - \gamma_2) \right) \quad (3.5)$$

To compute the optimal Lagrange coefficients $(\lambda_1^*, \lambda_2^*, \gamma_1^*, \gamma_2^*)$, the CBF inequality constraints (2.15), along with the optimal policies (3.5), are evaluated at the constraint boundaries.

$$\begin{pmatrix} I_m \\ -I_m \end{pmatrix} (G_v v^* + G_w w^*) = -\Delta G(x,w_f)$$
$$\begin{pmatrix} -I_m \\ I_m \end{pmatrix} H_w w^* = -\Delta H(x) \quad (3.6)$$

Substituting (3.5), yields

$$\begin{pmatrix} I_m \\ -I_m \end{pmatrix} \left( G_v R_v^{-1} G_v^T (\lambda_1 - \lambda_2) + G_w R_w^{-1} \left( G_w^T (\lambda_1 - \lambda_2) - H_w^T (\gamma_1 - \gamma_2) \right) \right) = 2\Delta G(x, w_f)$$

$$\begin{pmatrix} -I_m \\ I_m \end{pmatrix} H_w R_w^{-1} \left( G_w^T (\lambda_1 - \lambda_2) - H_w^T (\gamma_1 - \gamma_2) \right) = 2\Delta H(x) \tag{3.7}$$

Based on the complementary slackness conditions from (3.3), the system (3.7) can be decomposed into the four subsets of equations, each one representing a specific vector-boundary condition, with a single nonnegative Lagrange vector coefficient, and with rest of the coefficients set to zero.

$$\begin{aligned}
{[G_1 = 0]} &\Rightarrow \left[ \boxed{\lambda_1 \geq 0}, \ \lambda_2 = 0, \ \gamma_1 = 0, \ \gamma_2 = 0 \right] \Rightarrow \left[ \left( G_v R_v^{-1} G_v^T + G_w R_w^{-1} G_w^T \right) \lambda_1 = 2\Delta G_1(x, w_f) \right] \\
{[G_2 = 0]} &\Rightarrow \left[ \lambda_1 = 0, \ \boxed{\lambda_2 \geq 0}, \ \gamma_1 = 0, \ \gamma_2 = 0 \right] \Rightarrow \left[ \left( G_v R_v^{-1} G_v^T + G_w R_w^{-1} G_w^T \right) \lambda_2 = 2\Delta G_2(x, w_f) \right] \\
{[H_1 = 0]} &\Rightarrow \left[ \lambda_1 = 0, \ \lambda_2 = 0, \ \boxed{\gamma_1 \geq 0}, \ \gamma_2 = 0 \right] \Rightarrow \left[ H_w R_w^{-1} H_w^T \gamma_1 = 2\Delta H_1(x) \right] \\
{[H_2 = 0]} &\Rightarrow \left[ \lambda_1 = 0, \ \lambda_2 = 0, \ \gamma_1 = 0, \ \boxed{\gamma_2 \geq 0} \right] \Rightarrow \left[ H_w R_w^{-1} H_w^T \gamma_2 = 2\Delta H_2(x) \right]
\end{aligned} \tag{3.8}$$

Define two symmetric strictly positive definite matrices,

$$\begin{aligned}
R_{\lambda,\lambda} &= G_v R_v^{-1} G_v^T + G_w R_w^{-1} G_w^T \\
R_{\gamma,\gamma} &= H_w R_w^{-1} H_w^T
\end{aligned} \tag{3.9}$$

and rewrite (3.8) in matrix form.

$$\underbrace{\begin{pmatrix} R_{\lambda,\lambda} & 0_{m\times m} & 0_{m\times m} & 0_{m\times m} \\ 0_{m\times m} & R_{\lambda,\lambda} & 0_{m\times m} & 0_{m\times m} \\ 0_{m\times m} & 0_{m\times m} & R_{\gamma,\gamma} & 0_{m\times m} \\ 0_{m\times m} & 0_{m\times m} & 0_{m\times m} & R_{\gamma,\gamma} \end{pmatrix}}_{A} \underbrace{\begin{pmatrix} \lambda_1 \\ \lambda_2 \\ \gamma_1 \\ \gamma_2 \end{pmatrix}}_{X} = 2 \underbrace{\begin{pmatrix} \Delta G_1(x, w_f) \\ \Delta G_2(x, w_f) \\ \Delta H_1(x) \\ \Delta H_2(x) \end{pmatrix}}_{B} \tag{3.10}$$

The left-hand-side matrix $A$ in (3.10) is diagonal. From the PI control design formalism and from the relative-degree assumption for the limited output $z_{\lim}$, it follows that the $(m\times m)$ matrices $G_v$, $G_w$ and $H_w$ are nonsingular. Consequently, $R_{\lambda,\lambda}$ is nonsingular since it is a sum of two symmetric strictly positive definite matrices, and at the same time, $R_{\gamma,\gamma}$ is nonsingular by definition. Therefore, $A$ is invertible and the solution $X$ in (3.10) is well-defined.

Enforcing positivity requirement [5], gives the four optimal Lagrange vector coefficients,

$$\begin{pmatrix} \lambda_1^* \\ \lambda_2^* \\ \gamma_1^* \\ \gamma_2^* \end{pmatrix} = 2 \max \left( 0_{(4m)\times 1}, \begin{pmatrix} R_{\lambda,\lambda}^{-1} \Delta G_1(x, w_f) \\ R_{\lambda,\lambda}^{-1} \Delta G_2(x, w_f) \\ R_{\gamma,\gamma}^{-1} \Delta H_1(x) \\ R_{\gamma,\gamma}^{-1} \Delta H_2(x) \end{pmatrix} \right) \tag{3.11}$$

and the corresponding min-norm CBF-based control augmentation policy can be written explicitly.

$$\begin{aligned}
v^* &= -0.5 R_v^{-1} G_v^T \left( \lambda_1^* - \lambda_2^* \right) \\
w^* &= -0.5 R_w^{-1} \left( G_w^T \left( \lambda_1^* - \lambda_2^* \right) - H_w^T \left( \gamma_1^* - \gamma_2^* \right) \right)
\end{aligned} \tag{3.12}$$

With a proper selection of the weights $R_v$ and $R_w$ in the minimization cost $J(v,w)$ (2.15) and based on (3.9), the diagonal elements of $A$ (3.10) can be defined as scaled $(m\times m)$– identity matrices.

$$\left[ R_v = r_v G_v^T G_v, \quad R_w = r_w G_w^T G_w \right], \quad \text{where}: r_v, r_w > 0 \tag{3.13}$$

Then the optimal Lagrange coefficients (3.11) become,

$$\begin{pmatrix} \lambda_1^* \\ \lambda_2^* \\ \gamma_1^* \\ \gamma_2^* \end{pmatrix} = 2\max\left( 0_{(4m)\times 1}, \begin{pmatrix} \frac{1}{r_v}\Delta G_1(x, w_f) \\ \frac{1}{r_v}\Delta G_2(x, w_f) \\ \frac{1}{r_w}\Delta H_1(x) \\ \frac{1}{r_w}\Delta H_2(x) \end{pmatrix} \right) \tag{3.14}$$

and the two positive scalar constants $(r_v, r_w)$ can be used to tune the design by weighing the operational constraints appropriately. Overall, the selection of the minimization weights (3.13) gives a practical tuning guideline for the proposed min-norm CBF-based controller design.

For presentation clarity, the derived servo-control augmentation design is summarized in the table below.

| | |
|---|---|
| Open-loop LTI MIMO plant dynamics (1.1) | $\dot{x}_p = A_p x_p + B_p u$ |
| Regulated output (1.1) | $y_{reg} = C_{p\,reg} x_p + D_{p\,reg} u$ |
| Limited output (1.1) | $z_{\lim} = C_p x_p + D_p u$ |
| Output tracking error | $e_y = y_{reg} - y_{cmd}$ |
| Integrator tracking error dynamics with integrator AW (1.3) | $\dot{e}_{yI} = e_y + v$ |
| Extended open-loop system (1.4) | $\underbrace{\begin{pmatrix} \dot{e}_{yI} \\ \dot{x}_p \end{pmatrix}}_{\dot{x}} = \underbrace{\begin{pmatrix} 0_{m\times m} & C_{p\,reg} \\ 0_{n_p\times m} & A_p \end{pmatrix}}_{A} \underbrace{\begin{pmatrix} e_{yI} \\ x_p \end{pmatrix}}_{x} + \underbrace{\begin{pmatrix} D_{p\,reg} \\ B_p \end{pmatrix}}_{B} u + \underbrace{\begin{pmatrix} -I_m \\ 0 \end{pmatrix}}_{B_{cmd}}(y_{cmd}-v)$ |
| Commanded control input with baseline control augmentation term (1.5), (1.6) | $u^{cmd} = \underbrace{\underbrace{-K_I e_{yI}}_{u_I} \underbrace{-K_P x_p}_{u_P}}_{u_{bl}^{cmd}} + w$ |
| Servo-controller with position limits (1.8) | $u = \text{sat}_{u^{\min}}^{u^{\max}}(u^{cmd})$ |
| Filtered control augmentation (2.2) | $\dot{w}_f = K_w(w - w_f)$ |
| Control constraints (2.4) | $g(u_{bl}^{cmd}) = \begin{pmatrix} g_1(u_{bl}^{cmd}) \\ g_2(u_{bl}^{cmd}) \end{pmatrix} = \begin{pmatrix} u^{\min} - u_{bl}^{cmd} - w_f \\ u_{bl}^{cmd} - u^{\max} + w_f \end{pmatrix} \leq 0$ |
| Limited output constraints (2.4) | $h(x_p) = \begin{pmatrix} h_1(x_p) \\ h_2(x_p) \end{pmatrix} = \begin{pmatrix} z_{\lim}^{\min} - z_{\lim} \\ z_{\lim} - z_{\lim}^{\max} \end{pmatrix} \leq 0$ |
| State dependent constraints (2.6), (2.14) | $\Delta G(x, w_f) = \begin{pmatrix} \Delta G_1(x, w_f) \\ \Delta G_2(x, w_f) \end{pmatrix} = \begin{pmatrix} I_m \\ -I_m \end{pmatrix} (K_I e_y + K_P(A_p x_p + B_p u_{bl}^{cmd}) + K_w w_f) + \alpha_v g(u_{bl}^{cmd})$ <br> $\Delta H(x) = \begin{pmatrix} \Delta H_1(x) \\ \Delta H_2(x) \end{pmatrix} = \begin{cases} \begin{pmatrix} -I_m \\ I_m \end{pmatrix} (C_p x_p + D_p u_{bl}^{cmd}) + \begin{pmatrix} z_{\lim}^{\min} \\ -z_{\lim}^{\max} \end{pmatrix}, & \text{if } \det(D_p) \neq 0 \\ \begin{pmatrix} -I_m \\ I_m \end{pmatrix} C_p(A_p x_p + B_p u_{bl}^{cmd}) + \alpha_w h(x_p), & \text{if } \det(C_p B_p) \neq 0 \end{cases}$ |
| Auxiliary matrices (2.5), (2.14) | $G_v = K_I, \quad G_w = K_P B_p - K_w$ <br> $H_w = \begin{cases} D_p, & \text{if } \det(D_p) \neq 0 \\ C_p B_p, & \text{if } \det(C_p B_p) \neq 0 \end{cases}$ |
| Minimization cost weights (3.13) | $\left[ R_v = r_v G_v^T G_v, \quad R_w = r_w G_w^T G_w \right], \quad \text{where: } r_v, r_w > 0$ |
| Nonsingular symmetric positive definite matrices (3.9) | $R_{\lambda,\lambda} = G_v R_v^{-1} G_v^T + G_w R_w^{-1} G_w^T$ <br> $R_{\gamma,\gamma} = H_w R_w^{-1} H_w^T$ |
| Lagrange multiplier vector coefficients (3.11) | $\begin{pmatrix} \lambda_1^* \\ \lambda_2^* \\ \gamma_1^* \\ \gamma_2^* \end{pmatrix} = 2 \max \left( 0_{(4m) \times 1}, \begin{pmatrix} R_{\lambda,\lambda}^{-1} \Delta G_1(x, w_f) \\ R_{\lambda,\lambda}^{-1} \Delta G_2(x, w_f) \\ R_{\gamma,\gamma}^{-1} \Delta H_1(x) \\ R_{\gamma,\gamma}^{-1} \Delta H_2(x) \end{pmatrix} \right)$ |
| Min-norm optimal CBF-based control augmentation solution (3.12) | $v^* = -0.5 R_v^{-1} G_v^T (\lambda_1^* - \lambda_2^*)$ <br> $w^* = -0.5 R_w^{-1} \left( G_w^T (\lambda_1^* - \lambda_2^*) - H_w^T (\gamma_1^* - \gamma_2^*) \right)$ |

**Table 1 Min-norm Optimal CBF-based Servo-Control Augmentation Design Summary**

Figure 1 shows the system block-diagram with a baseline servo-controller and the min-norm optimal CBF-based state feedback augmentation.

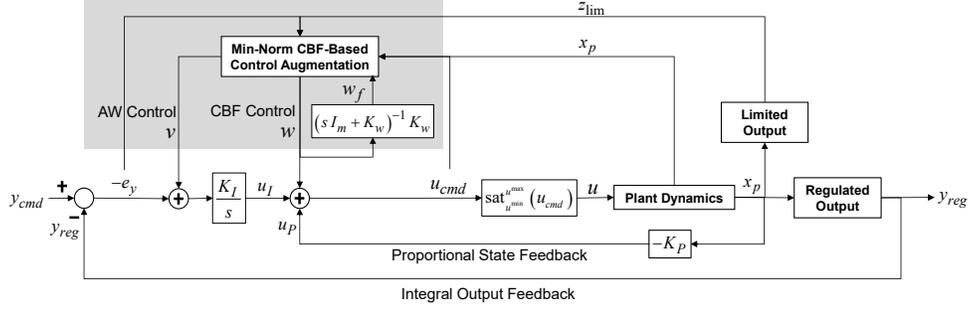

**Figure 1** Closed-loop system block-diagram

By design, the augmentation logic enforces soft constraints on the commanded control input, that is the closed-loop system dynamics would remain the same without the hard constraints imposed by the control saturation block. The latter is added to the diagram for practical purposes and in order to avoid errors due to numerical implementation / integration of the algorithm. Also note that the selected limited output is subject to soft constraints and requires no explicit hard saturation. In addition to soft-constrained control input and limited output signals, the augmentation solution adds anti-windup protection with respect to the controller integrator state components, keeping them uniformly bounded during control and/or output saturation events.

The derived min-norm optimal CBF-based servo-control augmentation solution (3.11), (3.12) represents a continuous state feedback linear control policy [8], [9] and as such, the corresponding closed-loop system stability and robustness properties can be directly analyzed using standard methods in linear systems [3].

## 4. Flight Control Design and Simulation Trade Study

Consider the roll-yaw dynamics representative of a mid-size aircraft, (see [3], Section 14.8, pp. 622–626).

$$\underbrace{\begin{pmatrix} \dot{\beta} \\ \dot{p}_s \\ \dot{r}_s \end{pmatrix}}_{\dot{x}_p} = \underbrace{\begin{pmatrix} \dfrac{Y_\beta}{V_0} & \dfrac{Y_{p_s}}{V_0} & \dfrac{Y_{r_s}}{V_0} - 1 \\ L_\beta & L_{p_s} & L_{r_s} \\ N_\beta & N_{p_s} & N_{r_s} \end{pmatrix}}_{A_p} \underbrace{\begin{pmatrix} \beta \\ p_s \\ r_s \end{pmatrix}}_{x_p} + \underbrace{\begin{pmatrix} \dfrac{Y_{\delta_{ail}}}{V_0} & \dfrac{Y_{\delta_{rud}}}{V_0} \\ L_{\delta_{ail}} & L_{\delta_{rud}} \\ N_{\delta_{ail}} & N_{\delta_{rud}} \end{pmatrix}}_{B_p} \underbrace{\begin{pmatrix} \delta_{ail} \\ \delta_{rud} \end{pmatrix}}_{u}$$

The system state $x_p$ includes the aircraft sideslip angle $\beta$ (rad), as well as the vehicle stability axis roll and yaw rates (rad/sec), $p_s$ and $r_s$. The control input $u$ is represented by the aileron and the rudder deflections (rad), $\delta_a$ and $\delta_r$. The regulated output of interest consists of the aircraft roll rate $p_s$ (rad/sec) and the lateral load factor $N_y$ (g-s), where $g = 32.174$ is the gravitational acceleration (ft/sec²).

$$y_{reg} = \begin{pmatrix} p_s & N_y \end{pmatrix}^T = \underbrace{\begin{pmatrix} 0 & 1 & 0 \\ \dfrac{Y_\beta}{g} & \dfrac{Y_{p_s}}{g} & \dfrac{Y_{r_s}}{g} \end{pmatrix}}_{C_{p\,reg}} x_p + \underbrace{\begin{pmatrix} 0 & 0 \\ \dfrac{Y_{\delta_{ail}}}{g} & \dfrac{Y_{\delta_{rud}}}{g} \end{pmatrix}}_{D_{p\,reg}} u = C_{p\,reg} x_p + D_{p\,reg} u$$

The aircraft model data are computed using numerical linearization with respect to a 1g-level flight trim (i.e., equilibrium) at the selected flight conditions.

$$V_0 = 717.17\left(\frac{ft}{\sec}\right), \quad Alt = 25000\,(ft), \quad \alpha = 4.5627\,(\deg)$$

$$A_p = \begin{pmatrix} -0.11794 & 0.00085 & -1.0001 \\ -7.0113 & -1.4492 & 0.22059 \\ 6.3035 & 0.06511 & -0.41172 \end{pmatrix}, \quad B_p = \begin{pmatrix} 0 & 0.015257 \\ -7.9662 & 2.6875 \\ 0.60926 & -2.3577 \end{pmatrix}$$

$$C_{p\,reg} = \begin{pmatrix} 0 & 1 & 0 \\ -2.6049 & 0.018724 & 0.067695 \end{pmatrix}, \quad D_{p\,reg} = \begin{pmatrix} 0 & 0 \\ 0 & 0.33698 \end{pmatrix}$$

A baseline LQR PI controller is designed without operational limits, using the integrated output tracking error dynamics,

$$\dot{e}_I = y_{reg} - y_{cmd} = \begin{pmatrix} p_s - p_{s\,cmd} \\ N_y - N_{y\,cmd} \end{pmatrix}$$

and the following LQR weights.

$$Q_{lqr} = \mathrm{diag}(1.025 \quad 1.0289 \quad 0 \quad 0 \quad 1.6021), \quad R_{lqr} = \mathrm{diag}(1 \quad 0.49129)$$

Figure 2 shows adequate closed-loop system tracking performance due to external step-input commands.

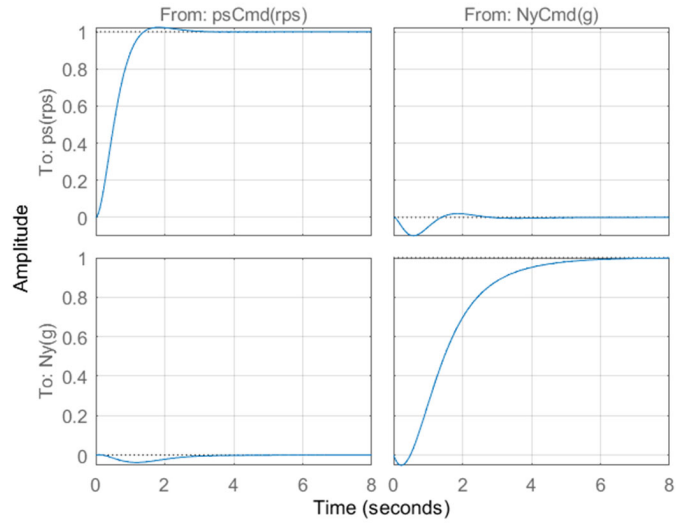

**Figure 2  Closed-loop system tracking performance with unconstrained baseline LQR PI controller**

Due to the tracking error integrators, dynamics of the two regulated outputs are almost decoupled. Figure 3 shows the LQR PI loop gains at the system input break-points, computed one at a time with and without an actuator model ("subsystems").

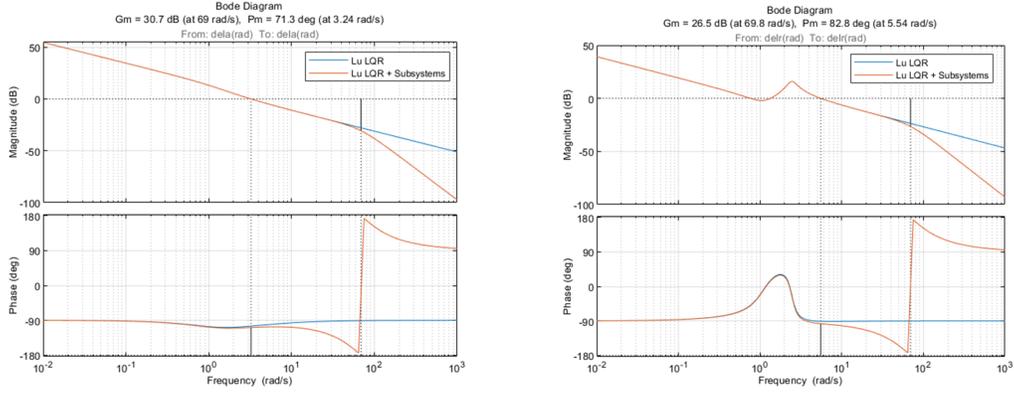

**Figure 3  Loop gains with unconstrained baseline LQR PI controller**

These data confirm satisfactory robustness and command tracking characteristics of the baseline controller, when it is operating without min/max limits.

Consider again the closed-loop system response using the unconstrained baseline LQR PI controller, which is tested with a series of step-input commands in $p_{s\,cmd}$ and a zero command in $N_{y\,cmd}$. In practical applications, such a test is very relevant and representative of demonstrating coordinated turn capabilities, as shown in Figure 4.

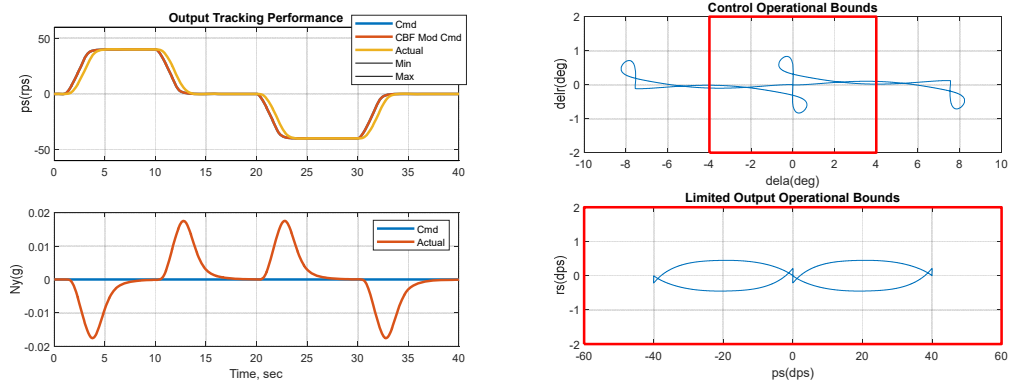

**Figure 4  Coordinated turn with unconstrained baseline LQR PI controller**

A min-norm CBF-based augmentation controller is designed using the guidelines from Table 1. Positive constants $\alpha_v$ and $\alpha_w$ are set to $10$. The filter matrix $K_w$ is set to four times the product of the LQR proportional gain $K_P$ and the system $B-$ matrix, $K_w = 4\left(K_P B_p\right)$.

For pure testing purposes, the aileron and the rudder position limits are set to $\left(\pm 4\,\mathrm{deg}\right)$ and $\left(\pm 2\,\mathrm{deg}\right)$, correspondingly, while the roll rate and the yaw rate limits are selected to be sufficiently large, $\left(\pm 60\,\mathrm{deg}\right)$ and $\left(\pm 2\,\mathrm{deg}\right)$. Selection of these operational limits allows to demonstrate efficiency of the nom-based CBF control augmentation with respect to control limits.

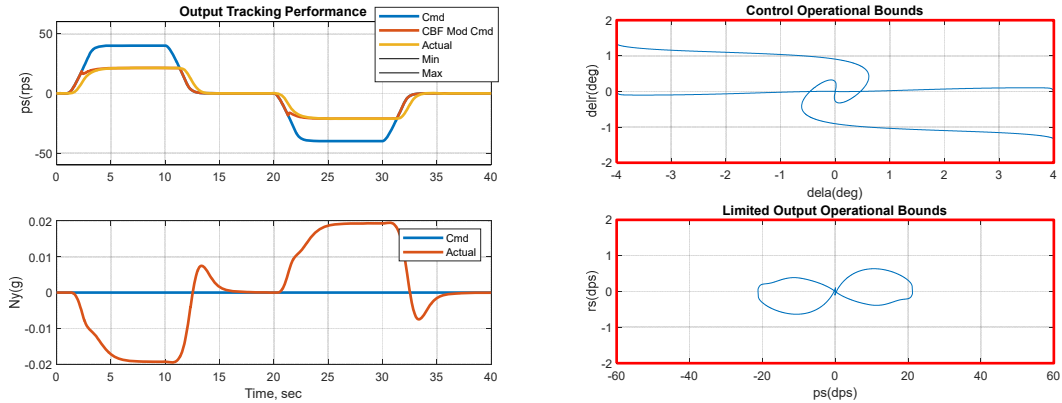

**Figure 5  Coordinated turn with constrained LQR PI controller and min-norm CBF-based control augmentation, in the presence of aileron limits**

As seen from the data, due to a large roll rate command, the aileron channel saturates (right upper phase plot) driving the achievable roll rate to become smaller than the command (left top plot). At the same time, the lateral acceleration (left bottom plot) remains small, indicating the desired turn coordination capabilities. Position and rate data for the same controller are shown in Figure 6.

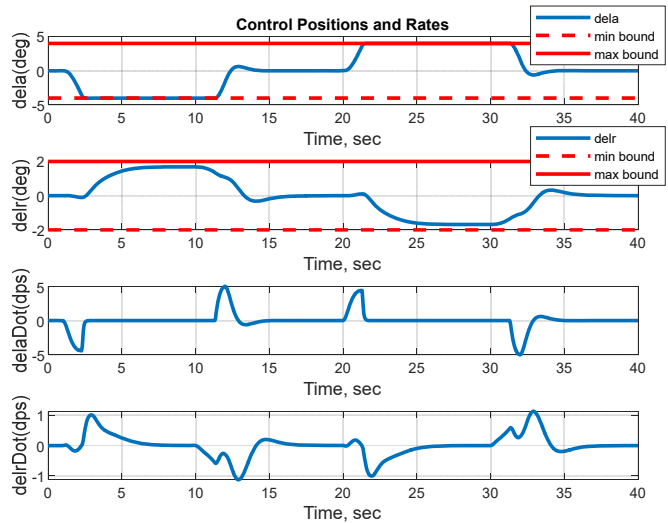

**Figure 6  Control positions and rates during coordinated turn with constrained LQR PI controller and min-norm CBF-based control augmentation**

In the next test, rudder limits are decreased to $(\pm 1\text{deg})$, in order to induce control position saturation in both control channels simultaneously, Figure 7.

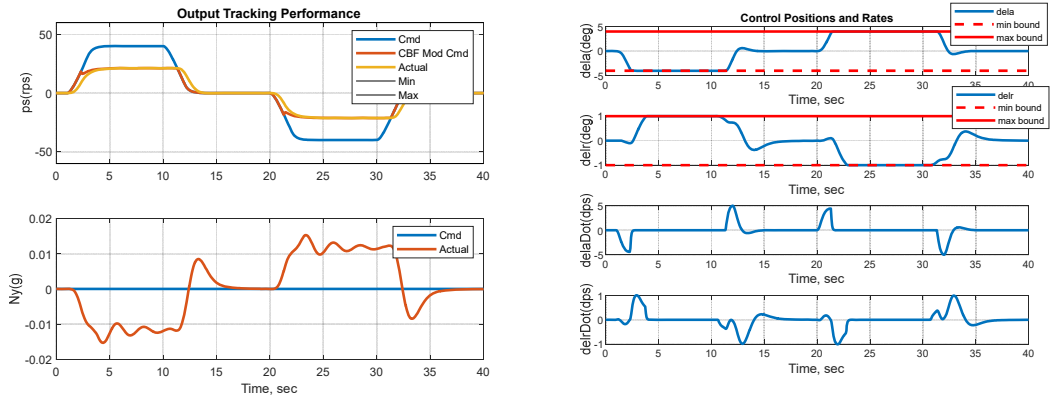

**Figure 7  Coordinated turn with two-channel saturated LQR PI controller and min-norm CBF-based control augmentation**

This test shows a clear benefit of the min-norm CBF-based augmentation logic. Even though both control channels are saturated, the controller is able to maintain closed-loop stability, retain turn coordination and continue tracking external commands to the extent possible. As expected per design, the commanded control values evolve within their designated limits, avoiding often undesirable "hard" saturation effects, Figure 8.

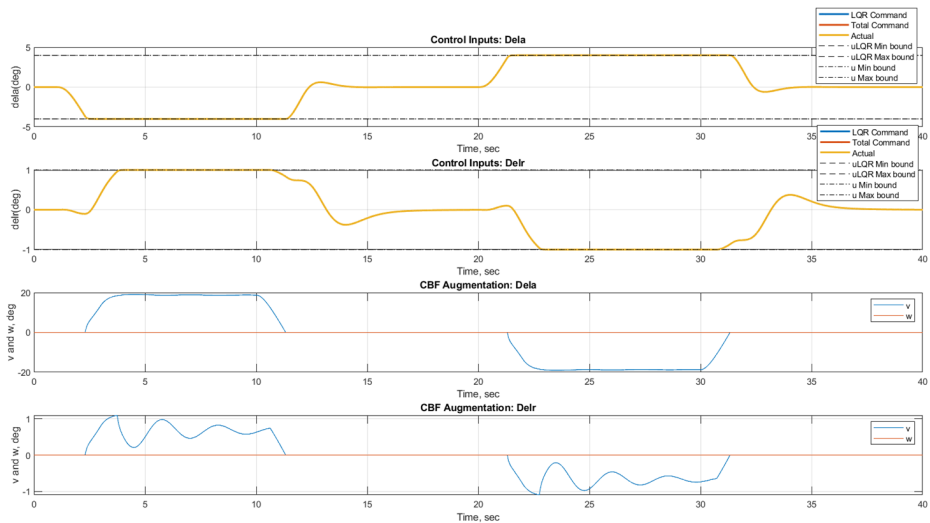

**Figure 8  Commanded and achieved control positions during coordinated turn with two-channel saturated LQR PI controller and min-norm CBF-based control augmentation**

The system states (plots 1, 2, 3) and the states of the control integrator (plot 4) are shown in Figure 9. In this case, the anti-windup signal $v$ enforces boundedness and smoothness of the controller integrators.

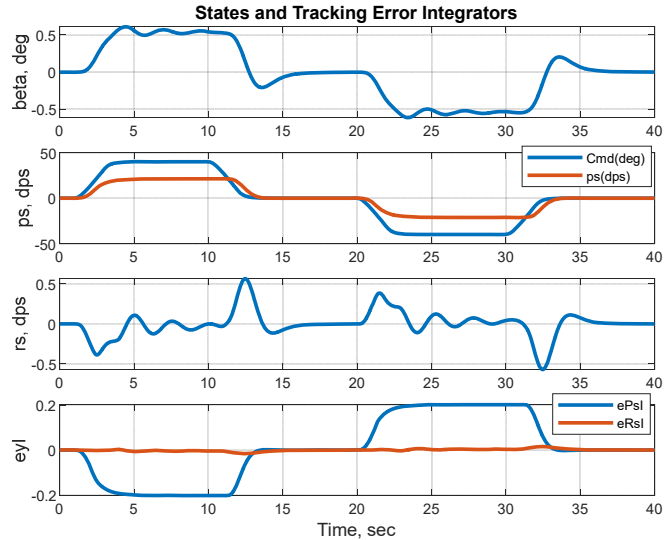

**Figure 9  System and controller states during coordinated turn with two-channel saturated LQR PI controller and min-norm CBF-based control augmentation**

In the next test, the rudder position bounds are increased back to $(\pm 2\,\text{deg})$ but the roll rate limits are decreased, $(\pm 18\,\text{deg})$, with respect to the min/max commanded rate, $(\pm 40\,\text{deg})$. Figure 10 shows simulation data with the baseline controller only, operating under the hard position saturation constraints.

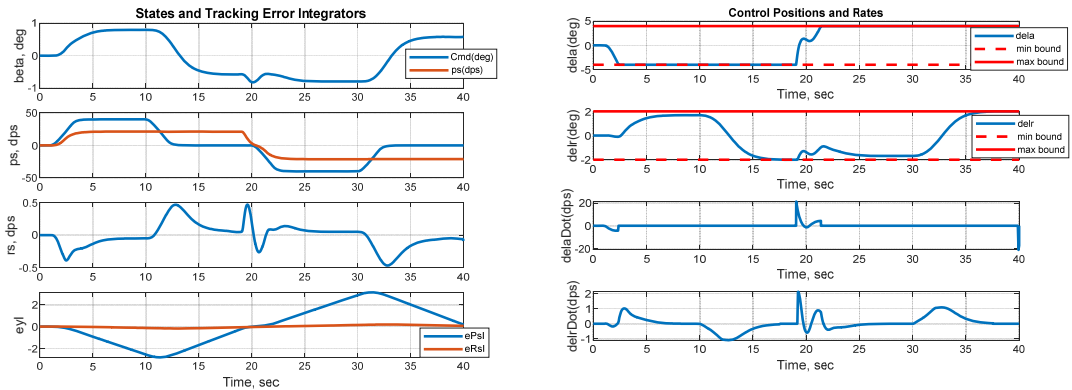

**Figure 10  Coordinated turn with baseline LQR PI controller and hard saturation constraints**

The system performance is clearly unacceptable. There is a very large time delay in the roll rate response (second plot, left) because the integrator state is winding-up (fourth plot, left) during the aileron saturation event (first plot, right). Turning the augmentation controller on, restores the baseline system performance that would be obtained without the hard control saturation in the loop and a properly reduced roll rate command, Figure 11.

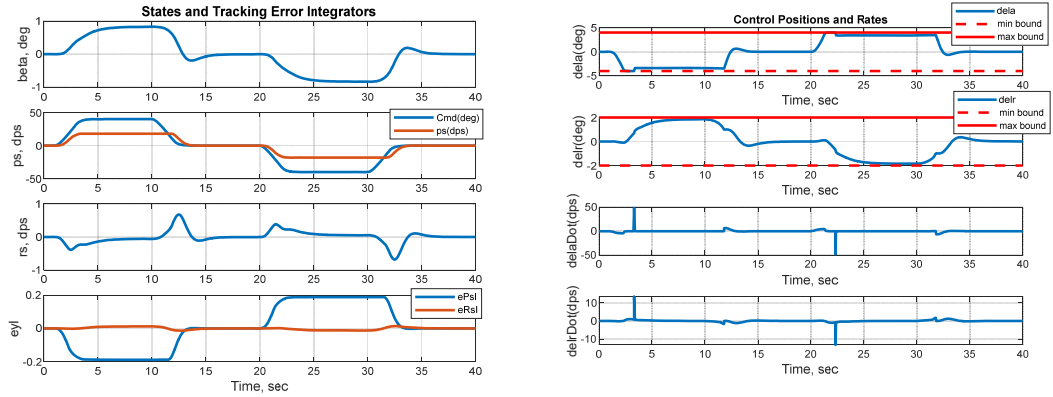

**Figure 11  Coordinated turn with (LQR PI + Min-norm CBF Augmentation) controller**

The system regulated output follows modified roll rate command, while the integrator state dynamics are bounded. There is no apparent time delay in the regulated output response. The min-norm CBF-based control augmentation calculates an adjustment to the commanded roll rate based on the desired limits only. It can also be verified that the aileron and roll rate min/max bounds are enforced as soft limits, Figure 12.

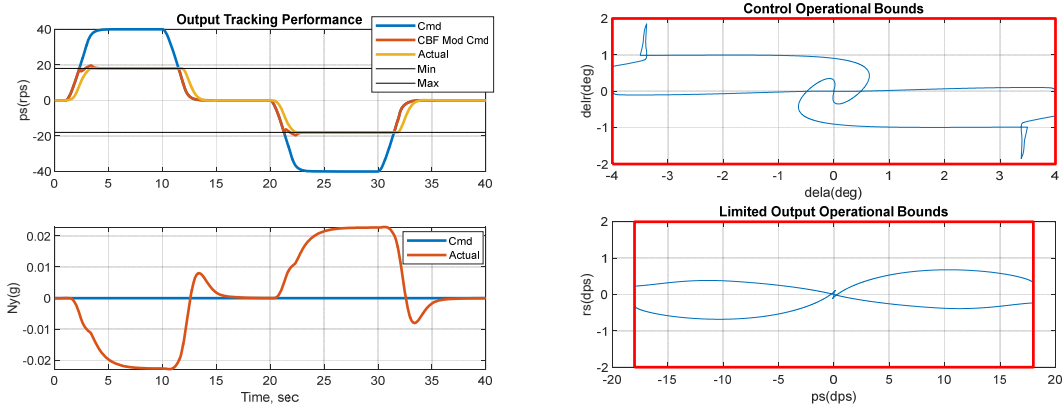

**Figure 12  Regulated output and controls for coordinated turn with (LQR PI + Min-norm CBF Augmentation)**

Overall, simulation test data show potential benefits of the developed control augmentation solution for flight critical control applications, such as aircraft primary flight control systems. Specifically, this technology can be used to design output and control limiters to enforce operational limits for aerial vehicles.

## 5. Conclusions

In this paper, a formal control augmentation design method is developed for MIMO LTI systems with a baseline PI servo-controller subject to box constraints that represent the desired operational limits imposed on the system control input and a selected output. The design is based on the Nagumo Theorem [6], the min-norm controllers [4] with QP optimization [5], and CBF-based methods [8], [9]. The developed solution provides anti-windup protection for the controller integrator state and it enforces soft min/max constraints on the total control command as well as on the selected output.